\documentclass[aps,superscriptaddress,prb,reprint,amssymb,showpacs]{revtex4-1}
\usepackage{amsmath,amssymb}
\usepackage{amsfonts,amsthm}
\usepackage{epsfig}
\usepackage{graphicx}
\usepackage{dcolumn}
\usepackage{color}
\usepackage{bm}
\begin{document}
\title{Quantum Rotors on the AB$_2$ Chain with Competing Interactions}
\author{Ant\^onio S. F. Ten\'orio}
\affiliation{Laborat\'orio de F\'{\i}sica Te\'orica e Computacional,
Departamento de F\'{\i}sica, Universidade Federal de Pernambuco, CEP 50670-901, 
Recife, Pernambuco, Brazil}
\author{R.~R. Montenegro-Filho}
\affiliation{Laborat\'orio de F\'{\i}sica Te\'orica e Computacional,
Departamento de F\'{\i}sica, Universidade Federal de Pernambuco, CEP 50670-901,
Recife, Pernambuco, Brazil}
\affiliation{Escola Polit\'ecnica de Pernambuco, Universidade de Pernambuco,
CEP 50750-470, Recife, Pernambuco, Brazil}
\author{M.~D. Coutinho-Filho}
\affiliation{Laborat\'orio de F\'{\i}sica Te\'orica e Computacional,
Departamento de F\'{\i}sica, Universidade Federal de Pernambuco, CEP 50670-901,
Recife, Pernambuco, Brazil}
\begin{abstract}
We present the ground state phase diagram of $q = 1/2$ quantum-rotor chains
with competing interactions (frustration) calculated through cluster variational mean
field approaches.  We consider two interaction patterns, named F$_1$ and F$_2$
models, between the quantum-rotor momentum and position operators, which follow 
exchange patterns of known one-dimensional spin-1/2 systems with a
ferrimagnetic state in their phase diagrams. The spin-1/2 F$_1$ model is known as
the diamond chain and is related to the azurite compound, while the spin-1/2
F$_2$ model was recently shown to present a frustration-induced condensation of
magnons. We provide a detailed comparison between the quantum-rotor phase
diagrams, in single- and multi-site mean-field approaches, and known results for
the spin-1/2 models, including exact diagonalization and density matrix renormalization
group data for these systems, as well as phase diagrams of the associated
classical models.
\end{abstract}


\maketitle

\section{Introduction}
The connection between O($n$) quantum-rotor (QR) and spin models on \mbox{$d$-dimensional lattices} has proved very useful 
in the context of phase transitions \cite{Auer, QPTsach}. About three decades ago, Hamer, Kogut and 
Susskind \cite{HKS} mapped \mbox{two-dimensional} O($n$) Heisenberg models \mbox{($n$ = 2, 3 and 4)} onto 
the corresponding [(1+1) spatial and time dimensions] nonlinear-sigma or QR models. The critical 
behavior was then inferred using strong-coupling expansion (high-temperature, 
\mbox{$g= kT/J  \rightarrow \infty$}, where $J$ is the spin coupling): a Kosterlitz-Thouless transition for the 
O(2) model and a prediction of critical points at zero coupling (Pad$\acute{e}$ continued) for both O(3) and 
O(4) models.
On the other hand, by mapping O(3) antiferromagnetic (AF) Heisenberg chains onto nonlinear sigma models in 
the semiclassical weak-coupling limit \mbox{($g = 2/S, S \rightarrow \infty)$}, Haldane \cite{Haldane} suggested 
that the ground state (GS) of chains with integral spins are gapped, while those with half-integral spins are gapless. 
Moreover, Shankar and Read \cite{Shankar} precisely clarified the distinction between gapped AF spin models,
characterized by the $\theta$ = 0 \textit{mod} 2$\pi$  topological term, and gapless models for which $\theta = \pi$ 
\textit{mod} 2$\pi$, including the connection of the latter with a Laplacian minimally coupled to the monopole 
potential\cite{Mono}. Following the above developments, Sachdev and Senthil \cite{QPTsach, ZPTsach} have presented 
a quite general mean-field and renormalization-group analysis of quantum phase transitions in magnets with the aid 
of \textit{generalized} QR models. In particular, they showed that,
under certain conditions, one can establish a mapping of double-layer antiferromagnets onto quantum rotors 
which sheds intuitive light on the way in which a QR can be used as an effective representation of a 
pair of antiferromagnetically coupled spins. Still in this context, a single-site MF approximation was used 
to study an effective Hamiltonian for spin-one bosons in an optical lattice in the presence of a magnetic 
field \cite{Imambekov} . Further, a QR description of the Mott-insulator transition in the 
Bose-Hubbard model within a functional-integral approach has also been elaborated in order to include particle
number fluctuation effects \cite{Polak}.
\par In this work we focus our attention on the study of the GS phase diagram of generalized quantum 
rotors on the frustrated AB$_2$ chain, which is depicted in \mbox{Fig. \ref{Figure01}}. The quantum rotors at each site 
are constrained, through sufficiently high values of the coupling $g$ (and the coupling $\alpha$ of the quartic term 
in the angular momentum), to mostly retain states with the minimum value of the angular momentum, i.e., 
$\ell = 1/2$, as the frustration parameter $J$ is varied, thus enabling us to make a direct comparison with 
the corresponding \textit{quantum} spin-1/2 AB$_2$ chains. We analyze two types of frustration, as illustrated in 
\mbox{Fig. \ref{Figure01}}, and try to interpret the derived phase diagrams in light of the ones of previous works 
on frustrated quantum spin-1/2 chains with the AB$_2$ topology \cite{TKS, Rene, TOHTK, frust1a, frust1b, cyclic}. Instead 
of attempting to formalize a specific (and probably rather complex) mapping between the rotor and the spin models, 
we have opted to treat the rotor chain numerically by using a cluster variational MF theory, 
supplemented with exact diagonalization (ED) via Lanczos algorithm\cite{Lanczos} and 
density matrix renormalization group (DMRG) \cite{DMRG} of finite-size spin-1/2 chains.  
\par With respect to spin systems, as a motivation on the experimental side, the compound azurite \cite{azurite} has 
been successfully explained by the distorted diamond chain model\cite{frust1b}, i.e., a system with three spin-1/2 
magnetic sites per unit cell and frustrated ferrimagnetic state. Also, along with the study on the effect of 
frustration\cite{TKS, Rene, TOHTK, frust1a, frust1b, cyclic}, for $J = 0$ this class of models shares its phenomenology and 
unit-cell topology with quasi-one-dimensional compounds, such as the line of trimer clusters present in copper 
phosphates\cite{Matsuda}  and the organic ferrimagnet PNNBNO\cite{Hosokoshi}. The modeling of the ferrimagnetic 
phase \cite{revisao} has been mainly undertaken in the context of other models such as Hubbard\cite{Hubb}, $t-J$\cite{tJ}, 
Ising\cite{ClHeis_I}, classical\cite{ClHeis_I} and quantum Heisenberg\cite{Heis}, including magnetic 
excitations \cite{magnexc1,magnexc2}, and the quantum spherical model\cite{Mario}. The occurrence of new phases induced by 
hole-doping of the electronic band\cite{hole} has also been carried out. 
\par This paper is organized as follows. In the next section we describe our QR model and 
numerical methods, and include in Appendix A a derivation of the matrix elements of the operators acting on 
the single-site Hilbert space represented by monopole harmonics. In \mbox{Sec. III} we use single-site variational 
MF theory to study the rotor models, for the two frustration cases, and discuss the shortcomings of 
this semiclassical approach. Then in \mbox{Sec. IV} we adopt a multi-site (two-unit cell) variational MF 
Hamiltonian, which provides a substantial improvement on the treatment of quantum fluctuation effects, particularly 
in connection with the case of frustrated interaction between quantum rotors on B sites at the same unit cell. Here we
treat the respective spin-1/2 systems by making use of ED and DMRG techniques in order to pave the way for a direct comparison
between rotors and spins. Finally we report our conclusions in \mbox{Sec. V}.

\section{Outline of the Theory  and Methods}

Quantum rotors can be classified according to their minimum angular momentum\cite{ZPTsach, Mono}: rotors with 
$q = 0$ have zero minimum angular momentum, which can be made to correspond to an even number of Heisenberg spins 
in an underlying spin model. On the other hand, we also have rotors with 
$q \neq 0$, where $q$ is chosen to have one of the values: 1/2, 1, 3/2, ..., as will be briefly clarified below. This 
is an extension of the former case, and quantum rotors with half-integer values of $q$ are duly suited to refer to an odd number 
of underlying spins-1/2 (at least one spin remains unpaired). We shall focus on $q = 1/2$-quantum rotors in view of the stated 
objective of comparing our results with those for the referred chains of spin-1/2 operators.
\par The three-component unit vector (operator)  
\mbox{$\hat{\bf{n}} = (\hat{n}_{x},\hat{n}_{y},\hat{n}_{z})$}, with $\hat{\bf{n}}^{2} =1$, describes 
the configuration space ($n$ space) of a rotor, while 
\mbox{$\hat{\bf{L}} =(\hat{L}_{x},\hat{L}_{y},\hat{L}_{z})$} stands for the canonically conjugate angular momenta. 
Setting $\hbar \equiv 1$, these quantities obey the commutation relations (operators at different sites commute):
\begin{eqnarray}
 {[}\hat{L}_{\mu},\hat{L}_{\nu}{]} & = & i\epsilon_{\mu\nu\lambda}\hat{L}_{\lambda}, \nonumber \\
 {[}\hat{L}_{\mu},\hat{n}_{\nu}{]} & = & i\epsilon_{\mu\nu\lambda}\hat{n}_{\lambda}, \nonumber \\
 {[}\hat{n}_{\mu},\hat{n}_{\nu}{]} & = & 0, 
 \label{comm}
\end{eqnarray}
where the Greek letters stand for the Cartesian components $x, y,z$ (summation over repeated indices is subtended and $\epsilon_{\mu\nu\lambda}$ is the 
Levi-Civita tensor) and 
\begin{equation}  
\hat{L}_{\mu}=-\epsilon_{\mu\nu\lambda}\hat{n}_{\nu} \left[ i\frac{\partial}{\partial\hat{n}_{\lambda}}
+ qA_{\lambda}(\hat{n}_{\mu}) \right ] -q\hat{n}_{\mu},
\label{Lmonopole}
\end{equation}  
which incorporates the effect of a Dirac monopole at the origin of $n$ space, whose vector potential may 
be conveniently chosen to satisfy
\begin{equation} 
\epsilon_{\mu\nu\lambda}\partial A_{\lambda}/\partial \hat{n}_{\nu} = \hat{n}_{\mu}.
\label{VecPot}
\end{equation}
\par The appropriate Hilbert space is made up of \textit{angular section} states for which the following are 
true \cite{Mono}:
$\hat{\bf{L}}^{2}|q,l,m> = l(l+1)|q,l,m>$; $\hat{L}_{z}|q,l,m>= m|q,l,m>$; and the usual ladder operators ($\hat{L}_{\pm} =\hat{L}_{x}\pm i\hat{L}_{y}$) satisfy $\hat{L}_{\pm}|q,l,m> = \sqrt{(l\mp m)(l\pm m+1)}|q,l,m \pm 1>$. Here  
$l=q, q+1, q+2,\ldots$, and  $m= -l, -l+1,\ldots, l$. The $|q,l,m>$ are the eigensections, also 
called \emph{monopole harmonics}. 
An important constraint follows immediately from \mbox{Eq.(\ref{Lmonopole})}: 
\begin{equation}
\hat{\bf{n}} \cdot \hat{\bf{L}} = -q.
\label{constraint} 
\end{equation}

\par
\begin{figure}
\begin{center}
\includegraphics*[width=0.4\textwidth,clip]{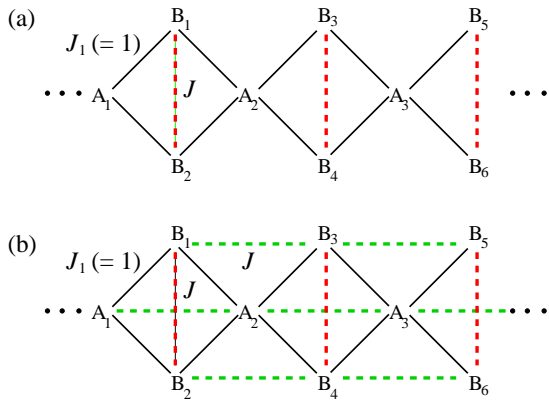} 
\caption{(Color online) Illustration of the QR chains with three rotors per unit cell \textit{i}: $A_i$, 
$B_{2i-1}$, and $B_{2i}$. Full lines indicate antiferromagnetic exchange couplings ($J_1 \equiv 1$) which give rise to 
the ferrimagnetic GS, while dashed lines represent exchange couplings ($J \geq 0$) which frustrate the magnetic 
order: (\textit{a}) frustration pattern F$_1$ and (\textit{b}) frustration  pattern F$_2$.}
\label{Figure01}
\end{center}
\end{figure}

 Thus, following Ref. 7, we shall consider the quite general frustrated O(3) QR Hamiltonian:
\begin{eqnarray}
                         & &\!\! \hat{H}_{R}=\frac{g}{2} \sum_{i}[(\hat{\bf{L}}_{i}^{2}+\alpha(\hat{\bf{L}}_{i}^{2})^2)] + \nonumber \\
                        & & \!\!\!\! \sum_{<ij>}[{\hat{\bf{n}}_{i}} \cdot \hat{\bf{n}}_{j}+{\hat{\bf{L}}_{i}}\cdot \hat{\bf{L}}_{j} 
			  + M({\hat{\bf{n}}_{i}} \cdot \hat{\bf{L}}_{j}+{\hat{\bf{n}}_{j}} \cdot \hat{\bf{L}}_{i})]+  \nonumber\\ 
                        & &\!\!\!\!\!\!\!\!\!\!\!\!\!  \sum_{(i,j) \in F_1 or F_2}\!\!\!\!\!\!\!\![J({\hat{\bf{n}}_{i}} \cdot \hat{\bf{n}}_{j}+{\hat{\bf{L}}_{i}} 
			  \cdot \hat{\bf{L}}_{j})+M ({\hat{\bf{n}}_{i}} \cdot \hat{\bf{L}}_{j}+{\hat{\bf{n}}_{j}} \cdot \hat{\bf{L}}_{i})],  
			  \label{Hr}   
\end{eqnarray}
where $g$ and $\alpha$ are positive local couplings, i.e., associated with rotors at each site $i$ of the AB$_2$ 
chain (the quartic term appears with the main objective of controlling contributions of high-energy states); in 
the second summation, $< ij >$ indexes nearest-neighbor couplings between rotors on distinct sublattices which, 
except for $M$, are all set to unity (see \mbox{Fig. \ref{Figure01}}); in the third summation $(i,j)$ indexes 
nearest-neighbor couplings between rotors on the same sublattice which, except for $M$, are set to $J(\geq 0)$. Here 
we shall study two frustration patterns, namely, $F_1$ and $F_2$. In $F_1$ only frustrated interactions ($J$ and $M$) between 
rotors at B sites of the same unit cell are present, as illustrated in Fig. \ref{Figure01}(a), whereas for F$_2$ we 
consider \textit{all} nearest-neighbor intra- and intercell interactions, as illustrated in Fig. \ref{Figure01}(b). 
In order to isolate the effect of the coupling $M$ in the two above-referred cases, we take either $M=0$ or $M=1$.
\par Before going on to the approaches described in \mbox{Sec. III} and \mbox{Sec. IV}, we emphasize the 
following features about the \textit{stability} of the numerical implementations carried out in this work.
\par In our simulations we have verified that we could work safely with a minimally reduced Hilbert space if 
the values of $g$ and $\alpha$ were set sufficiently large. In fact, the Hilbert space size and the value of $g$ 
and $\alpha$ determine the stability of our problem: for small space sizes (e.g., $\ell = 3/2$) and small 
values of $g$ and $\alpha$ (\textit{e.g.}, $g= \alpha = 0.1$), the system becomes completely unstable due to contributions of 
high-energy terms which cause the system to fluctuate beyond control.  
On the other hand, by choosing a small space size ($\ell = 3/2$), but a sufficiently large value of $g$, the 
system behaves quite stably. Therefore, in this work we shall use small space size, i.e., $\ell = 3/2$, 
associated with a large value of $g$, in order to make computations feasible and a close contact with 
spin-1/2 models.
\par We then start off by treating $\hat{H}_R$ by means of a variational MF theory based on the Bogoliubov
theorem \cite{Call,Yokoi}. Thus, the variational expression of the MF energy at $T = 0$ satisfies the inequality:
\begin{equation}
E_{mf} \leq E_0 + <\hat{H}_R - \hat{H}_{trial}>_0,
\label{Des-Bogol}
\end{equation}
where $E_0$ is the GS energy of the trial Hamiltonian - here denoted by $\hat{H}_{trial}$ - and the 
expectation value is taken with respect to its GS wavefunction. Eq. (\ref{Des-Bogol}) is 
then minimized with respect to its variational parameters: for chosen values of the frustration control 
parameter ($J$), minimization is carried out numerically through diagonalization of $\hat{H}_{trial}$ by way of the 
Lanczos algorithm \cite {Lanczos}, and then deploying a simplex procedure \cite{Simplex}.
\par For trial Hamiltonians we use both single-site and multi-site Hamiltonians, as described in \mbox{Sec. III} and \mbox{Sec. IV}, respectively.

\section{Quantum Rotors on the AB$_2$ Chain: Single-site Variational Mean-Field Approach on the Unit Cell}

\par As a first and straightforward application of the aforementioned variational MF theory, we postulate the 
following trial Hamiltonian, acting on one unit cell:
                  \begin{equation}
       \hat{H}_{trial} = \frac{g}{2} \sum_{i}[(\hat{\bf{L}}_{i}^{2}+\alpha(\hat{\bf{L}}_{i}^{2})^2) +
                        {\bf{N}}_{i} \cdot \hat{\bf{n}}_{i} + {\bf{h}}_{i} \cdot \hat{\bf{L}}_{i}],
                        \label{Hsingle}       
                        \end{equation}
where \mbox{$\mathbf{h} = ({h}_{x},{h}_{y},{h}_{z})$} and \mbox{$\mathbf{N} = ({N}_{x},{N}_{y},{N}_{z})$} are the variational
c-number fields and the subscript  $i$ goes over the sites $A_1$, $B_1$, and $B_2$.  
\par The GS wavefunction of $\hat{H}_{trial}$ and energy are given by 
$|{\Psi}_0 >  = |{\Psi}_0 >_{A_1} |{\Psi}_0 >_{B_1} |{\Psi}_0 >_{B_2}$ and 
$E_0 = \sum_{i}E_{0_i} =  E_{0_{A_1}} + E_{0_{B_1}} + E_{0_{B_2}}$,
where $E_{0_i}=  E_{0_i}(g, \alpha; N_i,h_i)$ represents the GS energy of the respective wavefunction, such
that for any pair of operators $\hat{\bf{X}}_{i}, \hat{\bf{X}}_{j}$, with $i\neq j$:
\begin{equation}
<{\Psi}_0|\hat{\bf{X}}_{i} \cdot \hat{\bf{X}}_{j}|{\Psi}_0>  =
< \hat{\bf{X}}_{i}>_0 \cdot <\hat{\bf{X}}_{j}>_0,
\label{Dotproduct} 
\end{equation}
We then get, for frustration F$_1$, the Bogoliubov inequality for the unit cell: 
$E_{mf}^{(F_1)}  \leq  E_1 + E_2 + E_3$, where the $E_\nu$ read:
\begin{eqnarray*}
E_1  =   \sum_{i}E_{0_i} - \sum_{i}({\bf{N}}_{i} \cdot <\hat{\bf{n}}_{i}>_0 + 
{\bf{h}}_{i} \cdot <\hat{\bf{L}}_{i}>_0);& &\\
 E_2 =  2 \sum_{i>j, j = A_1}[<\hat{\bf{n}}_{i}>_0 \cdot 
<\hat{\bf{n}}_{j}>_0 + <\hat{\bf{L}}_{i}>_0 \cdot <\hat{\bf{L}}_{j}>_0 & & \\ 
+M(<\hat{\bf{n}}_{i}>_0 \cdot <\hat{\bf{L}}_{j}>_0 + <\hat{\bf{n}}_{j}>_0 \cdot <\hat{\bf{L}}_{i}>_0)];
\end{eqnarray*}
and 
\begin{eqnarray*}
E_3 = J(<\hat{\bf{n}}_{B_1}>_0 \cdot <\hat{\bf{n}}_{B_2}>_0 + <\hat{\bf{L}}_{B_1}>_0 \cdot <\hat{\bf{L}}_{B_2}>_0)+& &\\ M(<\hat{\bf{n}}_{B_1}>_0 \cdot <\hat{\bf{L}}_{B_2}>_0 + <\hat{\bf{n}}_{B_2}>_0 \cdot<\hat{\bf{L}}_{B_1}>_0);
\end{eqnarray*}
the index $i$ ($j$) visits the sites of the unit cell, with the convention: $A_1 <B_1 < B_2$. For frustration 
F$_2$, a fourth term must be added to the Bogoliubov inequality: 
\begin{eqnarray*}
E_4 = 2\sum_{i} [J<\hat{\bf{n}}_{i}>_0 
\cdot <\hat{\bf{n}}_{i}>_0 + <\hat{\bf{L}}_{i}>_0 \cdot <\hat{\bf{L}}_{i}>_0 + & &\\ 
M(<\hat{\bf{n}}_{i}>_0 \cdot <\hat{\bf{L}}_{i}>_0 + <\hat{\bf{n}}_{i}>_0 \cdot <\hat{ \bf{L}}_{i}>_0)]. 
\end{eqnarray*}
The MF energy -- best evaluation of 
$E_{mf}^{(F_1)} \equiv E_{mf}^{(F_1)}(g, \alpha, J)$ or $E_{mf}^{(F_2)} \equiv E_{mf}^{(F_2)}(g, \alpha, J)$ --  is 
then obtained by performing the minimization with respect to variations of the fields ${\bf{N}}_{i}$ and 
${\bf{h}}_{i}$. 
\par To produce the results of this section, it sufficed to set $g = \alpha = 10$ and a space size determined 
by truncating the Hilbert space at $\ell = 3/2$. Further, we have focused only on those quantities that suffice to afford 
the relevant information needed for the 
proper interpretation of the problem at this level, i. e., the two-point MF momentum products, defined here 
through the products \mbox{$<\hat{\bf{L}}_{i}> \cdot <\hat{\bf{L}}_{j}>$}, where $i \neq j$ runs over the sites of 
the unit cell, and the MF energy. We thereby leave out the position- and momentum-position products, for they 
are redundant. This is due to the fact that $\hat{L}_\mu$ and  $\hat{n}_\mu$ have the same signature under all 
allowed symmetries for q$\geq 0$, and so their expectation values turn out to be proportional to each other on a 
given site \cite{ZPTsach}.

\begin{figure}
\begin{center}
\includegraphics*[width=0.47\textwidth,clip]{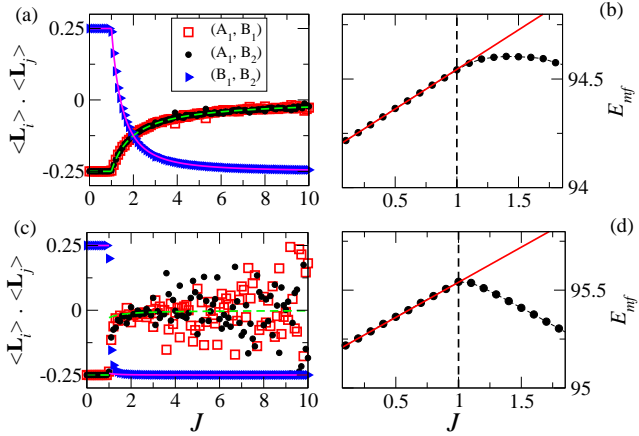}
\caption{(Color online) Frustration F$_1$. Two-point MF momentum products [(\textit{a}) $M \equiv 0$, (\textit{c}) $M \equiv 1$] 
between the indicated rotors and MF energy curve [(\textit{b})  $M \equiv 0$, (\textit {d}) $M \equiv 1$], where we have 
drawn straight (full) lines to show that at $J=1$ the system steers away from the linear regime that prevails for $J\leq 1$ and so 
a phase transition takes place. Full and dashed lines  in (\textit{a}) and (\textit{c}) indicate the results of the classical 
vector model. Dashed lines in (\textit{b}) and
(\textit {d}) are guides to the eye.}
\label{Figure02}
\end{center}
\end{figure}

\begin{figure}
\begin{center}
\includegraphics*[width=0.47\textwidth,clip]{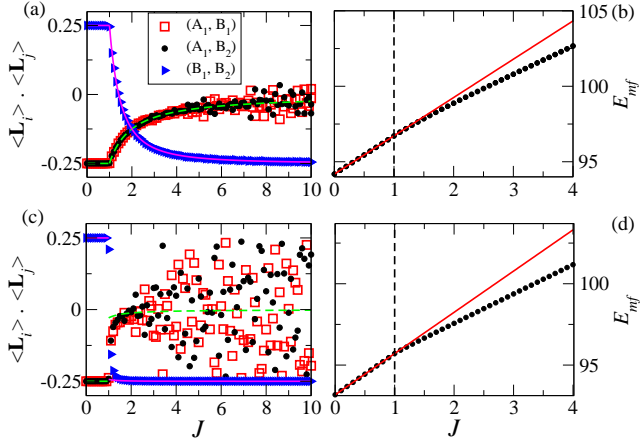}
\caption{(Color online) Frustration F$_2$. Same as in \mbox{Fig. \ref{Figure02}}.}
\label{Figure03}
\end{center}
\end{figure}

\par We then proceed to discuss the results in \mbox{Fig. \ref{Figure02}} (frustration F$_1$) and  
\mbox{Fig. \ref{Figure03}} (frustration F$_2$), which reveal some salient features. Firstly, we verified that 
\mbox{$<\hat{\bf{L}}_{i}>^2 = 0.25$}, with   $i=A_1, B_1, B_2$, independent of $J$. The momentum products 
show that, \textit{in all cases},  the system starts out with a magnetization plateau: 
\mbox{$<\hat{\bf{L}}_{B_1}>\cdot <\hat{\bf{L}}_{B_2}> = 0.25$} and 
\mbox{$<\hat{\bf{L}}_{A_1}>\cdot <\hat{\bf{L}}_{B_{1,2}}> = -0.25$}, which corresponds to the Lieb-Mattis\cite{Lieb} phase of 
the analogous spin-1/2 system, before undergoing a phase transition at \mbox{$J=1$}. This transition is of second order 
\mbox{($M \equiv 0$)}, as shown in \mbox{Fig. \ref{Figure02} (a)} and \mbox{Fig. \ref{Figure03} (a)}, and of first order 
\mbox{($M \equiv 1$)}, as shown in \mbox{Fig. \ref{Figure02} (c)}  and  \mbox{Fig. \ref{Figure03} (c)}.
\par In the first case ($M \equiv 0$), the system evolves continually (with the MF energy curve - \mbox{Fig. \ref{Figure02} (b)} 
and \mbox{Fig. \ref{Figure03} (b)} - smooth at the point $J=1$) to a stable phase where 
the momenta at A and B sites become uncorrelated, i.e., \mbox{$<\hat{\bf{L}}_{A_1}>\cdot <\hat{\bf{L}}_{B_{1,2}}> \approx 0$}, 
while the momenta at B sites tend to directly oppose each other with increasing $J$, forming a singlet-like configuration: 
\mbox{$<\hat{\bf{L}}_{B_1}>\cdot <\hat{\bf{L}}_{B_2}> \approx -0.25$}, for \mbox{$J \gg 1$}.
\par In the second case ($M \equiv 1$), the transition takes place quite abruptly, having undoubtedly first-order 
characteristics, and the system immediately accommodates into the stable singlet-like phase that we have just referred to. The 
MF energy curves of these first-order transitions at $J=1$ are shown in  \mbox{Fig. \ref{Figure02} (d)} and 
\mbox{Fig. \ref{Figure03} (d)}, where we notice that the cusp in the latter is less pronounced.
\par We notice further, that the products between the momenta at A and B sites display quite sizable fluctuations around \mbox{$<\hat{\bf{L}}_{A_1}>\cdot <\hat{\bf{L}}_{B_{1,2}}> \approx 0$}, as seen in \mbox{Fig. \ref{Figure02} (c)}  and  
\mbox{Fig. \ref{Figure03} (c)} ($M\neq 0$), and in lesser degree in \mbox{Fig. \ref{Figure03} (a)} for frustration $F_2$ and 
$M = 0$. The corresponding wide points occur pairwise and fairly symmetrically with respect to classical curves (see below) that 
represent the decoupling of the momenta at A and B sites, leaving the MF energy practically unaltered. In fact, with increasing $J$, the system becomes more prone to 
wandering through near-degenerate states, which give rise to these off points.  
\par The phase, for $J \gg 1$, with A sites  uncoupled and B sites with opposing momenta in a singlet-like
configuration is much like the dimer-monomer phase of the work by Takano, Kubo, and Sakamoto \cite{TKS}. We perceive, 
however, that important features in between those $J$ extremes of the phase diagram do not appear by way of 
this naive \textit{single-site} MF theory.

\begin{figure}
\begin{center}
\includegraphics*[width=0.28\textwidth,clip]{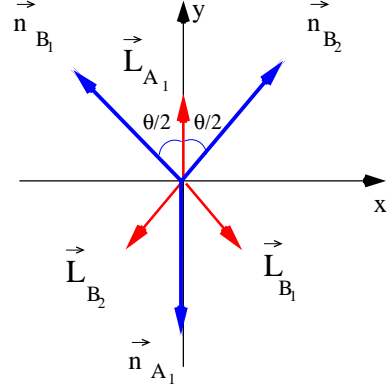}
\caption{(Color online) Classical vector configuration. The angle $\theta$ is the unique order parameter.} 
\label{Figure04}
\end{center}
\end{figure}

\par We now notice, through \mbox{Eq. (\ref{Dotproduct})}, that with the rotor momenta being fixed at ℓ
$\ell = 1/2$, we have $<L_i>^2 = 0.25$, for all sites $i$, independent of $J$, and so all products can only vary between the 
extremes  -0.25 and 0.25. Therefore, through this MF approach, we are led to envision the momenta on the unit cell of the 
AB$_2$ chain as classical vectors of constant magnitude, such as represented in \mbox{Fig. \ref{Figure04}}. We can thus provide 
a simple interpretation based on this configuration of classical vectors on the \textit{xy} plane (akin to the 
\textit{xy}-model). We then build our energy function for the configuration in \mbox{Fig. \ref{Figure04}} ($\alpha \equiv 0$) on a 
symmetric unit of the AB$_2$ chain centered on the A site. The classical constraints may be set as 
$|\mathbf{L}_{A_l}|=|\mathbf{L}_{B_1}|=|\mathbf{L}_{B_2}|\equiv 1/2$ and  $|\mathbf{n}_{A_l}|=|\mathbf{n}_{B_1}|=|\mathbf{n}_{B_2}|\equiv n$ (we take $n$ constant, which is about true for small $\ell$, as verified in the simulations, and whose value must be read off from the plots). We start off with the energy function for the frustration F$_1$ , taking into account the cases $M=0$ and $M=1$. So, for each case, up to a constant independent of $\theta$:
$E_{(M=0)}(\theta) = -2(1+4n^2) \cos\frac{\theta}{2} + \frac{J}{2}(1+4n^2)\cos\theta$ and 
$E_{(M=1)}(\theta) = -2(1- 4n+4n^2) \cos\frac{\theta}{2} - 2n \cos\theta + \frac{J}{2}(1+4n^2)\cos\theta$. Upon imposing the minimization conditions (relative to the unique parameter $\theta$), we obtain: (i)
for \mbox{$J<1$} we have $\theta = 0$, which holds for both $M=0$ and $M=1$; (ii) for $J > 1$, we have
$\theta \neq 0$, which in turn implies that $J =\frac{1}{\cos\frac{\theta}{2}}$, for $M=0$, while
$J =\frac{(1-2n)^2 +4n\cos\frac{\theta}{2}}{(1+4n^2)\cos\frac{\theta}{2}}$, for $M=1$.  The momentum 
products are accordingly given by: for \mbox{J $ < 1$}, $\mathbf{L}_{A_1} \cdot \mathbf{L}_{B_{1,2}} = -0.25$ and 
$\mathbf{L}_{B_1} \cdot \mathbf{L}_{B_2}= +0.25$,  for both $M=0$ and $M=1$. For J$ > 1$,
\begin{eqnarray}
\hspace{0.21in}\mathbf{L}_{A_1} \cdot \mathbf{L}_{B_{1,2}} & = &  -\frac{1}{4J}, \nonumber\\
\mathbf{L}_{B_1} \cdot \mathbf{L}_{B_2} & = &  \frac{1}{2J^2}-\frac{1}{4}, \hspace{0.1in} M=0;
\label{Cclassic2}
\end{eqnarray}
\begin{eqnarray}
\hspace{-0.21in} \mathbf{L}_{A_1} \cdot \mathbf{L}_{B_{1,2}} & = & -\frac {1}{4} (\frac{(1-2n)^2}{J(1+4n^2)-4n}),\nonumber \\
\hspace{-0.21in} \mathbf{L}_{B_1} \cdot \mathbf{L}_{B_2}  & = & \frac{1}{4}(\frac{2(1-2n)^4}{{[J(1+4n^2)-4n]}^2}\!-\!\!1),\hspace{0.05in}  M=1.     \label{Cclassic3}
\end{eqnarray}
\par With respect to frustration F$_2$, our present MF approach can only ``sense" a repetition of the configuration of 
\mbox{Fig. \ref{Figure04}}, in that  we get additional terms to the energy functions above 
that are independent of $\theta$ (and therefore vanish upon minimization), implying the same results for the dot products. 
\par This classical description fully accounts for the momentum
products in both frustration types for $M=0$, including the nature of the phase transition at $J=1$, as seen in
\mbox{Fig. \ref{Figure02}(a)} and \mbox{Fig. \ref{Figure03}(a)} through the matching fitting of the points of the 
numerical implementation for the rotors; for $M=1$, this interpretation confirms the first-order transition at $J=1$ and 
offers hints at the expected behavior of these momentum products, were it not for the off points, as can be seen in the 
diagrams of \mbox{Fig. \ref{Figure02}(c)} and \mbox{Fig. \ref{Figure03}(c)}. In  Eq. (\ref{Cclassic3}) we have used 
\mbox{$n = 0.34$}, that can be read off from plots of $<\hat{\bf{n}}^2>$, which were not explicitly presented in this work.
\par In the following section we try a \textit{more elaborate} MF technique on a double-cell structure, 
as a way to circumvent \mbox{Eq. (\ref{Dotproduct}}), as well as to get a direct evaluation of the intercell two-point 
correlations. For simplicity, we restrict ourselves to $M=0$.

\section{Quantum Rotors on the AB$_2$ Chain: Double-Cell Variational Mean-Field Approach}

Differently from the approach presented in Sec. III, we build our trial Hamiltonian acting on the \emph{global} space formed by 
the six sites of the double-cell structure made up of two contiguous unit cells, such as showed 
in \mbox{Fig. \ref{Figure01}}, i.e., we build one \emph{six-site} trial Hamiltonian acting, say, on the sites  
A$_1$, A$_2$, B$_1$, B$_2$, B$_3$, and B$_4$. In order to achieve that, we assigned to each site its own 
\emph{local} vector subspace, and we then constructed our global space by forming the tensor product of 
these subspaces in one chosen order. In order to simplify the equations below, when necessary, the 
\emph{local} operator acting on the QR located at site A$_1$, for instance, was denoted 
by $\hat{\bf{X}}_{A_1}$, which may refer to either operator $\hat{\bf{L}}$ or  operator $\hat{\bf{n}}$. 
\par We now write down the trial Hamiltonian acting on a given double-cell structure for frustration F$_1$: 
 \begin{equation}
 \hat{H}_{trial} = \hat{H}_{trial}^{(1)} + \hat{H}_{trial}^{(2)} + \hat{H}_{trial}^{(3)}. 
 \label{Htrial-ext}
 \end{equation}
 The first term (the kinetic energy term plus effective fields) is given by
                       \begin{equation}
 \hat{H}_{trial}^{(1)} =  \frac{g}{2}\sum_{i}[(\hat{\bf{L}}_{i})^{2}+\alpha((\hat{\bf{L}}_{i})^{2})^2 + 
                        {\bf{N}}_i \cdot \hat{\bf{n}}_i + {\bf{h}}_i \cdot \hat{\bf{L}}_i], 
\label{Htrial-ext1}
                        \end{equation}
where the index $i$ goes over the sites A$_1$, A$_2$, B$_1$, B$_2$, B$_3$, and B$_4$; $\mathbf{N}_i$ and
$\mathbf{h}_i$ being the effective fields (variational c-numbers) due to the rest of the system (which plays the role 
of a bath), and acting on each site $i$ of the double-cell cluster. The next term (first neighbors or bonds) reads
		        \begin{eqnarray} 
 \hat{H}_{trial}^{(2)} = \sum_{\hat{\bf{X}} = \hat{\bf{n}},\hat{\bf{L}}} [(\hat{\bf{X}}_{A_1} + 
                      \hat{\bf{X}}_{A_2})\cdot (\hat{\bf{X}}_{B_1} +
                      \hat{\bf{X}}_{B_2}) & & \nonumber \\
                      + \hat{\bf{X}}_{A_2} \cdot (\hat{\bf{X}}_{B_3} +\hat{\bf{X}}_{B_4})],& &
\label{Htrial-ext2}
		      	\end{eqnarray}
while the last one (frustration interaction) is written as
			\begin{equation} 
\hat{H}_{trial}^{(3)} = J\sum_{\hat{\bf{X}} = \hat{\bf{n}},\hat{\bf{L}}}
                     [(\hat{\bf{X}}_{B_1} \cdot  \hat{\bf{X}}_{B_2}) + (\hat{\bf{X}}_{B_3} \cdot 
                     \hat{\bf{X}}_{B_4})],
\label{Htrial3-ext}       
                        \end{equation}
where in the first term we opted to use explicit operators.
\par For frustration F$_2$, the following term (intercell frustration interaction) must be added to Eq. (\ref{Htrial-ext}):

			\begin{equation}			
\hat{H}_{trial}^{(4)} = J\sum_{\hat{\bf{X}} = \hat{\bf{n}},\hat{\bf{L}}} 
                                               [(\hat{\bf{X}}_{A_1} \cdot \hat{\bf{X}}_{A_2} + 
                                              \hat{\bf{X}}_{B_1}\cdot \hat{\bf{X}}_{B_3} +
					      \hat{\bf{X}}_{B_2}\cdot \hat{\bf{X}}_{B_4})]. 
\label{Htrial4-ext}	       
			\end{equation}

\par The direct application of \mbox{Eq. (\ref{Des-Bogol})} yields for frustration F$_1$ the 
following expression for the \textit{double-cell} variational MF energy $E^{(F1)}_{mf} \equiv E^{(F1)}_{mf}(g,\alpha,J)$,
where the equals sign implies that minimization with respect to the variational fields has already been carried out:
\begin{eqnarray}
E^{(F1)}_{mf} & = & E_0 +\sum_{\hat{\bf{X}} = \hat{\bf{n}},\hat{\bf{L}}}<\hat{\bf{X}}_{A_1}>_0 \cdot 
( <\hat{\bf{X}}_{B_3}>_0 + <\hat{\bf{X}}_{B_4}>_0) \nonumber \\
            &   & -\sum_{i}({\bf{N}}_i \cdot \hat{\bf{n}}_i + {\bf{h}}_i \cdot \hat{\bf{L}}_i), 
\label{Emf-multi}
\end{eqnarray}
where, $E_0$ represents the GS energy of $\hat{H}_{trial}$ and, as before, the index $i$ visits the sites A$_1$, A$_2$, B$_1$, B$_2$, B$_3$, and B$_4$.
\par For frustration F$_2$, we get analogously $E^{(F2)}_{mf} = E^{(F2)}_{mf}(g,\alpha,J)$: 			
                         \begin{eqnarray}			 
E_{mf}^{({F_2})} & = & E_0 + \sum_{\hat{\bf{X}} = \hat{\bf{n}},\hat{\bf{L}}} 
                       <\hat{\bf{X}}_{A_1}> \cdot (<\hat{\bf{X}}_{B_3}> + <\hat{\bf{X}}_{B_4}>) \nonumber \\  
	          &   & +2J\sum_{\hat{\bf{X}} = \hat{\bf{n}},\hat{\bf{L}}}  
	                (<\hat{\bf{X}}_{A_1}> \cdot <\hat{\bf{X}}_{A_2}> + \nonumber\\			
	          &   & <\hat{\bf{X}}_{B_1}> \cdot <\hat{\bf{X}}_{B_3}> +
		        <\hat{\bf{X}}_{B_2}> \cdot <\hat{\bf{X}}_{B_4}>)  \nonumber \\
                  &   &- \sum_{i}({\bf{N}}_i \cdot 
		        \hat{\bf{n}}_i + {\bf{h}}_i \cdot \hat{\bf{L}}_i).
\label{Emf-multi2}       
			\end{eqnarray}
\par Now, a given eigenfunction of $\hat{H}_{trial}$ may not necessarily be a tensor product of the eigenfunctions of the 
respective site subspaces as was the case in \mbox{Sec. III}, so that, for example, 
\begin{equation}
<{\Psi}_0|\hat{\bf{X}}_{A_1} \cdot \hat{\bf{X}}_{B_2}|{\Psi}_0>  \neq  <{\Psi}_0|\hat{\bf{X}}_{A_1}|{\Psi}_0> \cdot
 <{\Psi}_0|\hat{\bf{X}}_{B_2}|{\Psi}_0>,
\label{Exp2}
\end{equation}
where $|{\Psi}_0>$ designates the GS wavefunction of $H_{trial}$.  This is an important aspect in 
our approach, which differs from the standard MF result given by \mbox{Eq. (\ref{Dotproduct})}. Thus, in principle, taking 
advantage of the available capability of diagonalizing more complex operators (trial Hamiltonians), we can produce more 
reliable cluster variational MF theories.
\par The dimension of the global space is $d^6$, where $d$ is the dimension of the local subspace, so due to computational
implementability, this fact prompted us to limit the size of the Hilbert space by deploying rotors with maximum $\ellℓ = 3/2$ . 
\par An observation about the value of g is in order. In our approach, for frustration $F_1$,  when we had set 
$g = 10$ as in the preceding section, we verified that the momentum correlation 
$ <\mathbf{\hat{L}}_{A_1} \cdot \mathbf{\hat{L}}_{A_2}>$ remains pegged at 0.25, even 
after the transition at $J = 2$, which turns out not to be true (see below). Therefore, in this section, we resorted to a higher value of $g$ 
($g = 1000$), which inhibited more strongly the appearance of disturbing states; however one should notice that if much greater 
values of $g$ are employed, the kinetic energy becomes overwhelmingly dominant, so that small changes in the correlations 
tend to go unnoticed.
\par As before, we concentrated on the relevant quantities that can provide the information needed for the 
physical interpretation of the problem: namely, \textit{the mean-field energy}, the \textit{expectation value of 
the total angular momentum}, and the \textit{momentum correlations}. The expectation value of the total 
angular momentum per unit cell was calculated according to the formula
\mbox{$|<\mathbf{L}>|^2 = 1/2\sum_{\mu} <L_\mu^2>$}, where $L_\mu$ ($\mu = x,y,z$) is the respective resultant component
(component sum over all the six sites of the double-cell structure).
 
\subsection{Frustration F$_1$}

\begin{figure}
\begin{center}
\includegraphics*[width=0.43\textwidth,clip]{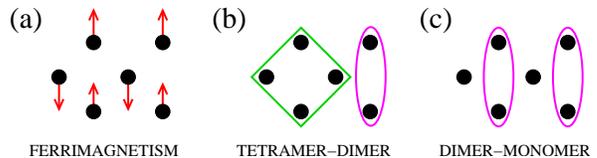}
\caption{(Color online) Illustration of the ground-states found for the spin-1/2 diamond chain\cite{TKS} as $J$ is increased
from $0$. (\textit{a}) The ferrimagnetic (FERRI) state. (\textit{b}) The 
tetramer-dimer (TD) state, where rectangles represent singlet tetramers and ellipses singlet dimers. (\textit{c}) 
The dimer-monomer (DM) state. There are two first-order phase transitions: at $J=0.909$ (FERRI/TD) and  
$J= 2$ (TD/DM).}
\label{Figure05}
\end{center}
\end{figure}

In order to facilitate comparison between QR results and those for the spin-1/2 counterpart with the same 
type of frustration, we present in \mbox{Fig. \ref{Figure05}} the phases obtained for the spin-1/2 diamond chain\cite{TKS}. 
The Lieb-Mattis\cite{Lieb} ferrimagnetic phase (FERRI) appears when $J < 0.909$ ($J$ is also used to indicate the frustration control parameter 
for the spin system). In the tetramer-dimer (TD) phase, which ensues when $0.909 < J<2$, the state is precisely the 
regular array of singlet tetramers (the closed loop encompasses four spins, in which the B sites form a triplet pair, and the 
spins on the A sites oppose those on the B sites, so that zero total spin takes place), and singlet dimers (two spins 
within the elliptical contour) as shown in \mbox{Fig. \ref{Figure05} (b)}. Finally, the dimer-monomer (DM) state 
is shown in \mbox{Fig. \ref{Figure05} (c)} and sets in when $J>2$; it is composed of the regular array of singlet 
dimers and free spins, and vanishing total spin is also expected. Because of the free spins, the DM state is 
macroscopically $2^{N/3}$-–fold degenerate for a chain with $N$ sites. Furthermore, both transitions are 
of first order \cite{TOHTK}.

\begin{figure}
\begin{center}
\includegraphics*[width=0.43\textwidth,clip]{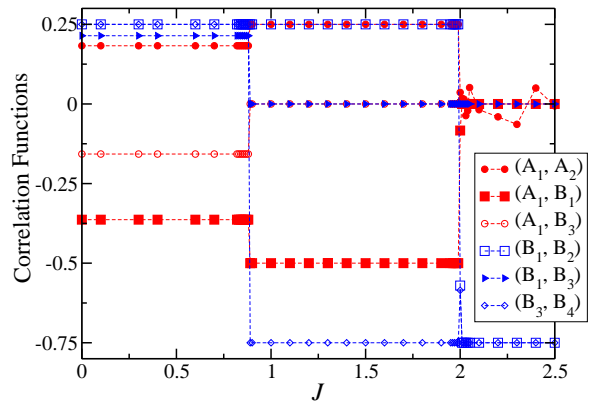}
\caption{(Color online) Spin-1/2 diamond chain: ED results for the correlation functions between spins at 
a central cluster of a system with 28 sites. Dashed lines are guides to the eye.}
\label{Figure06}
\end{center}
\end{figure}

\begin{figure}
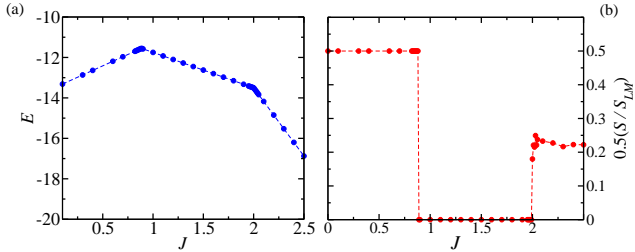

\begin{center}
\includegraphics*[width=0.23\textwidth,clip]{fig7a.eps}
\includegraphics*[width=0.23\textwidth,clip]{fig7b.eps}
\caption{(Color online) Spin-1/2 diamond chain: ED results for the (\textit{a}) average ground-state energy
and (\textit{b}) rescaled total spin of a system with 28 sites. Phase transitions occur at $J = 0.88$ and $J = 2.0$, both of 
first order. Dashed lines are guides to the eye.}
\label{Figure07}
\end{center}
\end{figure}

\par In order to allow a direct comparison with our MF results for quantum rotors, we have solved the spin-1/2 diamond chain 
(AB$_2$ chain with frustration between spins at sites B of the same unit cell) for 
sizes up to 28 sites, using the Lanczos ED procedure with open boundary conditions. The results are displayed as follows: the relevant correlations are represented 
by the curves plotted in \mbox{Fig. \ref{Figure06}}; in \mbox{Fig. \ref{Figure07} (a) and (b)} we plotted respectively the 
energy and total-spin curves (normalized by the Lieb-Mattis value \cite{Lieb}). 
\par Examination of the correlation plots show clear correspondence with the phases exhibited in 
\mbox{Fig. \ref{Figure05}}. The phase FERRI is characterized by the following correlations: 
\mbox{$ <\hat{\mathbf{S}}_{B_1} \cdot \hat{\mathbf{S}}_{B_2}> = <\hat{\mathbf{S}}_{B_3} \cdot \hat{\mathbf{S}}_{B_4}> = 
0.25$}, $ <\hat{\mathbf{S}}_{B_1} \cdot \hat{\mathbf{S}}_{B_3}> = 0.21$, 
$<\hat{\mathbf{S}}_{A_1} \cdot \hat{\mathbf{S}}_{A_2}> = 0.18$, 
$<\hat{\mathbf{S}}_{A_1} \cdot \hat{\mathbf{S}}_{B_3}> = -0.15$, 
$<\hat{\mathbf{S}}_{A_1} \cdot \hat{\mathbf{S}}_{B_1}> = -0.36$.
 The total spin per unit cell in \mbox{Fig. \ref{Figure07} (b)} shows the Lieb-Mattis value of 0.5 
throughout. The transition to the intermediate phase TD then occurs at 
$J = 0.88$, very close to the estimated value for the infinite chain \cite{TKS}: $J=0.909$.  We note that in this phase the 
chain breaks up into smaller units \mbox{- tetramers and dimers -} and quantum fluctuations within each unit do not affect 
the spin correlations. Hence the correlations are just those calculated for the TD configuration of spins 
in \mbox{Fig. \ref{Figure05} (b)}: 
\mbox{$ <\hat{\mathbf{S}}_{A_1} \cdot \hat{\mathbf{S}}_{A_2}> = <\hat{\mathbf{S}}_{B_1} \cdot \hat{\mathbf{S}}_{B_2}> = 0.25$} (triplets), 
\mbox{$ <\hat{\mathbf{S}}_{B_3} \cdot \hat{\mathbf{S}}_{B_4}> = -0.75$} (singlets), \mbox{$ <\hat{\mathbf{S}}_{A_1} \cdot 
\hat{\mathbf{S}}_{B_1}> = -0.5$}, 
the other correlations being zero. With increasing $J$ though, quantum fluctuations become strong enough 
to disrupt the tetramer unit and a new phase transition to DM phase happens at $J = 2$, this point being independent of  
size because of the chain breakup. In this phase correlation \mbox{$ <\hat{\mathbf{S}}_{A_1} \cdot \hat{\mathbf{S}}_{A_2}>$} has
varying nonzero values and does not vanish, as expected in the thermodynamic limit, due to finite size effects. 
On the other hand, the B spins, which
are interlocked in singlet units, are totally unaffected. This phase, depicted in 
\mbox{Fig. \ref{Figure05} (c)}, shows the final chain breakup as the tetramer gives way to two monomer units 
and another dimer, clearly indicated by the correlations in \mbox{Fig. \ref{Figure06}} ($J\geq 2$). The energy curve 
in \mbox{Fig. \ref{Figure07} (a)}  exhibits cusps at the transition points, typical of a first-order nature, also verified through 
the discontinuities of the correlations at these points. The total spin per unit cell in  \mbox{Fig. \ref{Figure07} (b)}
corroborates the above phase description; however in the last phase the apparent nonzero value is a finite-size effect. 
\par
\begin{figure}
\begin{center}
\includegraphics*[width=0.43\textwidth,clip]{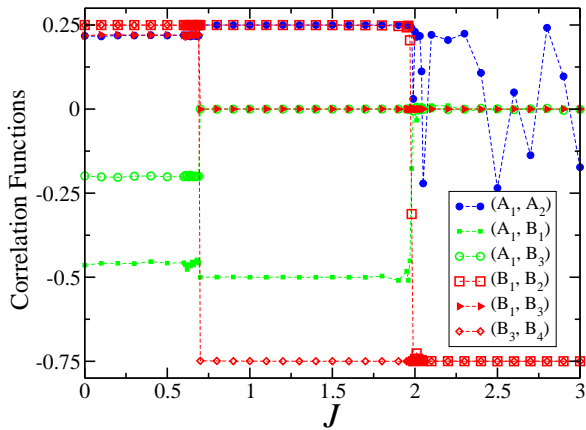}
\caption{(Color online) QR momentum correlations calculated by using the double-cell variational MF approach for frustration 
F$_1$. One notices the phase sequence FERRI$\leftrightarrow$TD$\leftrightarrow$DM, with first-order transitions at $J=0.68$ and $J=2$. Dashed lines 
are guides to the eye.}
\label{Figure08}
\end{center}
\end{figure}

\begin{figure}
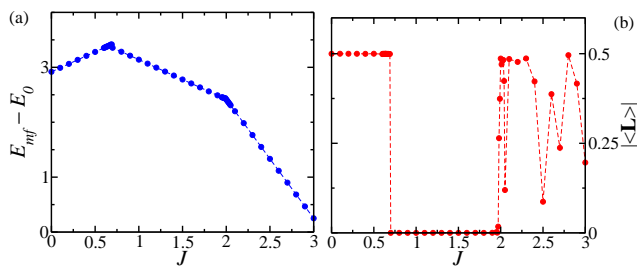

\begin{center}
\includegraphics*[width=0.23\textwidth,clip]{fig9a.eps}
\includegraphics*[width=0.23\textwidth,clip]{fig9b.eps}
\caption{(Color online) Quantum rotors by using the double-cell variational MF results for frustration 
F$_1$: (\textit{a}) the energy plot ($E_0 = 1962$) shows cusps at the first-order transition points $J = 0.68$ and 
$J = 2.0$; (\textit{b}) expectation value of the total angular momentum per unit cell. Dashed lines are guides to the eye.}
\label{Figure09}
\end{center}
\end{figure}

 Finally, getting down to the QR AB$_2$ chain, we display our variational MF numerical results in 
\mbox{Fig. \ref{Figure08}} (momentum correlations) and \mbox{Fig. \ref{Figure09}} (energy and total angular momentum) and  we 
proceed to a comparative examination with respect to the preceding spin results. A blow-by-blow confrontation of the correlations 
in both \mbox{Fig. \ref{Figure06}} and \mbox{Fig. \ref{Figure08}} shows that the double-cell variational MF approach 
is able to reproduce the three phases exhibited in Ref. 10, namely, the FERRI, TD, and DM phases. In the FERRI phase, 
quantum fluctuations appear to be equally important, causing the same correlations to deviate somewhat from calculated results 
for stiff momenta. A closer examination shows that, up to two decimal digits, we have same correlations for $<\hat{\mathbf{L}}_{B_1}\cdot\hat{\mathbf{L}}_{B_2}> = 
<\hat{\mathbf{L}}_{B_3}\cdot\hat{\mathbf{L}}_{B_4}> = 0.25$, but slightly different correlations, namely:
$<\hat{\mathbf{L}}_{A_1}\cdot\hat{\mathbf{L}}_{A_2}> = 0.22$, $<\hat{\mathbf{L}}_{A_1}\cdot\hat{\mathbf{L}}_{B_1}> = 0.46$, 
$<\hat{\mathbf{L}}_{A_1}\cdot\hat{\mathbf{L}}_{B_3}> = 0.20$, and $<\hat{\mathbf{L}}_{B_1}\cdot\hat{\mathbf{L}}_{B_3}> = 0.22$, 
for the rotor system, which should be compared with
$<\hat{\mathbf{S}}_{A_1}\cdot\hat{\mathbf{S}}_{A_2}> = 0.18$, $<\hat{\mathbf{S}}_{A_1}\cdot\hat{\mathbf{S}}_{B_1}> = 0.36$, 
$<\hat{\mathbf{S}}_{A_1}\cdot\hat{\mathbf{S}}_{B_3}> = 0.16$, and $<\hat{\mathbf{S}}_{B_1}\cdot\hat{\mathbf{S}}_{B_3}> = 0.21$, 
for the spin system. Phase transitions occur 
at $J = 0.68$ and $J = 2$, evidently of first order; in the first transition we have a lesser value ($J = 0.68$) than the 
Lanczos result for the 28-site spin-1/2 chain ($J = 0.88$), and that of Ref. 10 ($J = 0.909$). The momentum and spin correlations
match one another, respectively, in both phases: in the DM phase, correlation $<\hat{\mathbf{L}}_{A_1}\cdot\hat{\mathbf{L}}_{A_2}>$
shows also an erratic behavior similar to its spin counterpart, in other words, finite size effects are also at play. The minor differences in correlations in the FERRI phase, as well 
as in the first transition point do not constitute a fundamental discrepancy between the respective phase diagrams, which are endowed with the same topological features.
In \mbox{Fig. \ref{Figure09} (a)} the cusps in the mean-field energy at $J =0.68$ and $J = 2$ also bespeak the 
occurrence of these first-order transitions. Comparing the total momentum in \mbox{Fig. \ref{Figure09} (b)} with the total spin 
in \mbox{Fig. \ref{Figure07} (b)}, we observe similar results for the Lanczos predictions for the spin model, including finite-size 
effects in the last phase.

\par
\begin{figure}
\begin{center}
\includegraphics*[width=0.30\textwidth,clip]{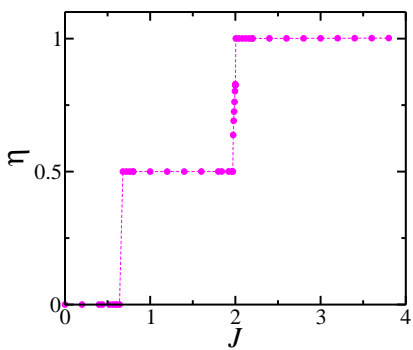}
\caption{(Color online) Quantum rotors with frustration F$_1$: average singlet density per unit cell for 
the momenta of the B sites at the same unit cell. One can make out the three phases: FERRI, TD, and DM, as well as 
pertinent transitions. Dashed lines are guides to the eye.}
\label{Figure10}
\end{center}
\end{figure}

 It is instructing to study the QR system regarding the average singlet density per unit cell of the
B momenta \cite{Rene}, which in our case (\textit{double-cell} cluster) is calculated directly using 
\begin{equation}
<\eta > = \frac{1}{4}-\frac{1}{2}(<\hat{\bf{L}}_{B_1} \cdot \hat{\bf{L}}_{B_2} > + <\hat{\bf{L}}_{B_3} \cdot \hat{\bf{L}}_{B_4} >), 
\label{singlet-density-F1}
\end{equation}
and which is displayed in \mbox{Fig. \ref{Figure10}}. These results permit a direct comparison with the phase diagram of 
\mbox{Fig. \ref{Figure05}}, as far as the buildup of singlet pairs out of B momenta is concerned. As is the case for spins, size effects are not important here, so that one perceives that the number of singlets is very clearly a quantized 
quantity within each phase.

\subsection{Frustration F$_2$}

Before getting down to quantum rotors, we describe succinctly existent results\cite{Rene} for the spin-1/2 AB$_2$ chain with
the same frustration F$_2$ pattern. The rich phase diagram of the model was studied 
through DMRG, exact diagonalization, and a hard-core boson model. The phase diagram thus obtained presented
three transition points. The first one is continuous and occurs at $J=0.34$ between the Lieb-Mattis ferrimagnetic phase (F1) and 
a ferrimagnetic phase (F2) characterized by the condensation of the singlet component of spins at sites B$_1$ and B$_2$ of the same 
unit cell, with transverse critical antiferromagnetic correlations.  At $J = 0.445$, a first-order
transition to a phase characterized by spiral and predominantly AF correlations (\textit{singlet spiral}) takes place. The number 
of singlets in the lattice is quantized before this transition, but is a continuous quantity afterwards, and can be envisioned 
by measuring the singlet density. 
Further, a continuous chain-ladder decoupling transition at 
$J = 0.91$ is observed. Above this value, the A spins present critical AF correlations following the asymptotic behavior 
observed in a linear chain, with power-law decay, while the ladder of B spins are short-range correlated with a finite correlation length, whose value is $J$-dependent, and nears the two-legged-ladder configuration (\textit{decoupled chain ladder}).

\begin{figure}
\begin{center}
\includegraphics*[width=0.45\textwidth,clip]{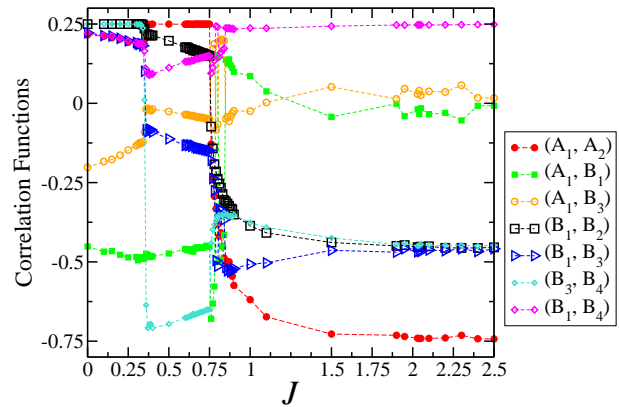}
\caption{(Color online) QR momentum correlations calculated by using the double-cell variational MF approach for frustration 
F$_2$. One can distinguish three major phases: FERRI, CANTED, and the decoupled AF chain ladder system, with transitions 
occurring around $J = 0.35$ and  $J = 0.75$. Dashed lines are guides to the eye.}
\label{Figure11}
\end{center}
\end{figure}

\begin{figure}
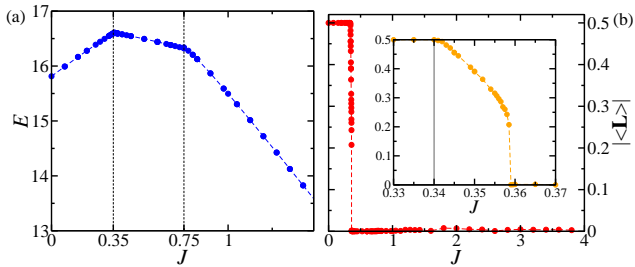

\begin{center}
\includegraphics*[width=0.23\textwidth,clip]{fig12a.eps}
\includegraphics*[width=0.23\textwidth,clip]{fig12b.eps}
\caption{(Color online) Quantum rotors by using the double-cell variational MF results for frustration 
F$_2$: (\textit{a}) energy and (\textit{b}) expectation value of the total angular momentum per unit cell. The inset shows details of the phase transition around $J=0.35$. Dashed lines are guides to the eye.}
\label{Figure12}
\end{center}
\end{figure}

\par
\begin{figure}
\begin{center}
\includegraphics*[width=0.47\textwidth,clip]{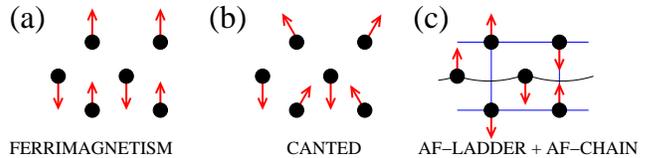}
\caption{(Color online) Illustration of the major QR ground states for frustration F$_2$: (\textit{a}) FERRI; (\textit{b}) CANTED;
(\textit{c}) AF, which is composed of two decoupled 1D systems: a linear chain (A sites) and the two-legged ladder (B sites).}
\label{Figure13}
\end{center}
\end{figure}

 As far as quantum rotors are concerned, the examination of the momentum correlations in 
\mbox{Fig. \ref{Figure11}} reveals that the system starts out with the FERRI phase which is the counterpart of phase F1 of the 
spin system studied in Ref. 12. The double-cell variational MF energy plotted in \mbox{Fig. \ref{Figure12} (a)} exhibits 
a pattern quite similar to that of frustration F$_1$, shown in \mbox{Fig. \ref{Figure07} (a)}.
But resorting to \mbox{Fig. \ref{Figure11}} with the help of \mbox{Fig. \ref{Figure12} (b)} (total average momentum per unit cell),
we can clear up the picture: in fact, at $J= 0.34$ its reasonable to think that a second-order transition takes place giving rise 
to a narrow transient phase that corresponds to the phase F2 (condensation of singlet components of spins at sites B of the same
unit cell) for spin system and is best 
visualized through the inset in the latter figure, which shows the behavior of the total angular momentum. In this phase
the momenta of the A sites keep their ferromagnetic configuration ($<\hat{\mathbf{L}}_{A_1}\cdot\hat{\mathbf{L}}_{A_2}> = 0.25$)
while the B momenta conform to a magnetic canted configuration. A first-order 
transition follows at $J \approx 0.36$ to a new state that should correspond to the phase \textit{singlet spiral} of the respective spin system. With respect to the momenta at 
the A sites, the ferromagnetic configuration also prevails 
in this phase. The total angular momentum of the A sublattice exactly counterbalances that of the B sublattice, so that, as 
it happens for the spin system, a vanishing expectation value of the total angular momentum (spin) per unit cell 
occurs. Furthermore, upon inspecting the B correlations {Fig. \ref{Figure11}}, this phase appears here to have also a semiclassical
canted configuration, hence the name CANTED that we use to designate this QR phase together with the previous one.  In the same manner for the spin 
system \cite{Rene}, 
the additional intercell interactions produce nonquantized values of the B momenta (see, for example, 
correlation $<\hat{\mathbf{L}}_{B_1}\cdot\hat{\mathbf{L}}_{B_3}>$). The nonquantization verified in the spin system, which is 
a coherent superposition of singlet and triplet configurations, may rather 
be seen as manifestation of the symmetry-breaking of the invariance of the Hamiltonian under interchange of the B sites in the 
same cell brought about by the additional frustration. This is also clearly verified in the QR system. In the absence of the 
additional frustration, as is the the case for frustration $F_1$ (\mbox{Fig. \ref{Figure01} (a)}), this symmetry stays unscathed, 
so that there is no singlet-triplet superposition: we have either a singlet or a triplet configuration per cell, but never 
both simultaneously, which was already the case for both spin and QR systems.  Finally, as seen in \mbox{Fig. \ref{Figure11}}, at 
$J= 0.75$, quantum fluctuations bring the sudden decoupling of the chain through another phase transition with 
first-order characteristics (in the spin system the transition is second-order), and the system settles into an 
antiferromagnetic (AF) phase, also marked by a vanishing expectation value of the total angular momentum per unit cell, as shown 
in \mbox{Fig. \ref{Figure12} (b)}. In this phase the frustrated AB$_2$ chain splits into two decoupled chains, namely, an AF 
linear chain ($<\hat{\mathbf{L}}_{A_1}\cdot\hat{\mathbf{L}}_{A_2}> = -0.75$) and an AF two-legged ladder 
($<\hat{\mathbf{L}}_{B_1}\cdot\hat{\mathbf{L}}_{B_2}> = <\hat{\mathbf{L}}_{B_1}\cdot\hat{\mathbf{L}}_{B_3}> = 
<\hat{\mathbf{L}}_{B_1}\cdot\hat{\mathbf{L}}_{B_3}> = -0.5$  and $<\hat{\mathbf{L}}_{B_1}\cdot\hat{\mathbf{L}}_{B_4}> = 0.25$); 
the decoupling is seen through $<\hat{\mathbf{L}}_{A_1}\cdot\hat{\mathbf{L}}_{B_{3,4}}> \cong 0$. In \mbox{Fig. \ref{Figure13}}, 
we show a pictorial representation of the the three major phases FERRI, CANTED and AF. With respect to this AF phase, our QR 
simulations evidently shed no light onto the criticality and short-rangedness of the linear and two-legged ladder chains, respectively. 
This phase corresponds to the \textit{decoupled chain-ladder} system, which in turn has a vanishing total spin. 
The first-order transition at $J=0.75$ may rather be seen as a manifestation 
of finite-size effects of our two-cell approach: the absence of many intermediate states preclude a smooth transition. 

\par
\begin{figure}
\begin{center}
\includegraphics*[width=0.30\textwidth,clip]{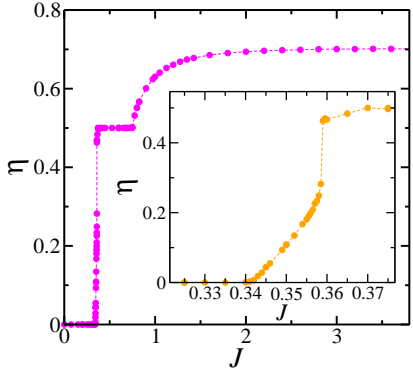}
\caption{(Color online) Quantum rotors with frustration F$_2$: average singlet density per unit cell for 
the momentum correlations at B sites along the same rung of the ladder.}
\label{Figure14}
\end{center}
\end{figure}

 The average singlet density for this frustration pattern is shown in \mbox{Fig. \ref{Figure14}}: the singlet number 
is quantized (except for the narrow interval around $J = 0.36$, until the frustration reaches the value 
$J = 0.75$, wherefrom the singlet number goes on nonquantized. We see that the QR system exhibits a four-phase pattern
quite similar to that of the respective spin system, with the nature of all but the last phase transitions being similar 
in both systems. With respect to singlet quantization, we find agreement in the first and last phases (where singlet densities 0 and 0.7 
are observed); in the intermediate phases no match is observed and again we impute this naturally to finite size effects of 
our two-cell approach, which hinder a discrete one-by-one singlet condensation. Also, because of the additional 
intercell frustration, it was not possible to form isolated singlet configurations as was the case for frustration F$_1$. 
\par
\begin{figure}
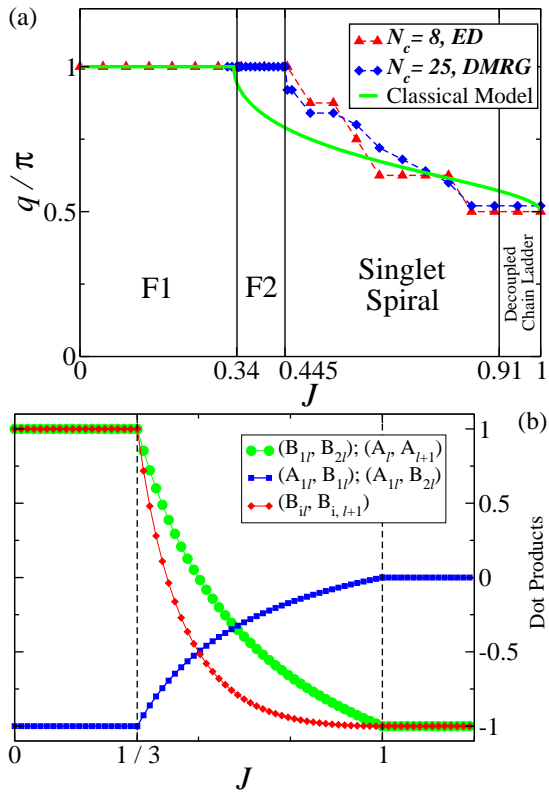

\begin{center}
\includegraphics*[width=0.4\textwidth,clip]{fig15a.eps}
\includegraphics*[width=0.4\textwidth,clip]{fig15b.eps}
\caption{(Color online) (\textit{a}) Pitch angle for the quantum spin-1/2 model calculated through ED,
DMRG, and for the minimum energy configuration of the classical vector model with two order parameters: $q$ (pitch angle) 
and $\theta$ (canting angle). The transition points estimated in Ref. 12 are indicated. (\textit{b}) Momentum dot products
($i= 1, 2$, and $l$ denotes the unit cell) in the minimum energy configuration of the classical vector model.}
\label{Figure15}
\end{center}
\end{figure}

 In what follows, we provide a more detailed comparison between QR MF results and the spin-1/2 chain, in its quantum
and classical versions. The phase diagram initially described of the spin-1/2 chain from Ref. 12 is summarized in 
\mbox{Fig. \ref{Figure15} (a)}. The spiral phase can be exposed in a clear fashion through the pitch angle $q$ obtained from 
the magnetic structure factor defined as 
\begin{equation}
F(q) = \frac{1}{2N_c}\sum_{j,k}<\hat{\mathbf{S}}_j \cdot \hat{\mathbf{S}}_k> e^{iq(j-k)}, 
\end{equation}
with $q=2\pi n/(2N_c)$, for $n= 0, 1, 2, \ldots
2N_c-1$, and $\mathbf{S}_j = \mathbf{A}_{(j+1)/2}$, if $j$ is odd, while $\mathbf{S}_j = \mathbf{B}_{1,j/2} + \mathbf{B}_{2,j/2}$,
if $j$ is even, and here we are labeling the sites in a more convenient way: A$_1$, B$_{1l}$, and B$_{2l}$ just denote
the sites A$_1$, B$_1$, and B$_2$ of the $lth$ unit cell. In the Lieb-Mattis phase the ferrimagnetic order is indicated by a sharp peak at $q = \pi$ 
(a period-2 configuration); while in the decoupled phase, in which a period-4 structure is observed 
(see \mbox{Fig. \ref{Figure13}}), there is a peak at $q=\pi/2$. These two situations are magnetic configurations commensurate with
the lattice, while the spiral phase is indicated by a peak at a value of $q$ between $q=\pi/2$ and $q=\pi$. In  
\mbox{Fig. \ref{Figure15} (a)} we display the behavior of $q$ as a function of $J$ for finite systems calculated through ED 
and DMRG. Finite size effects lead to a little shift in the transition point from the spiral phase to the 
decoupled phase, even though $q$ can be clearly used to mark the spiral phase. 

\par Motivated by these results, we consider the classical model in the space spanned by two parameters (in approach of Sec. III, 
one single parameter sufficed to explain the results): 
a canting angle $\theta$ between the B momenta at the same cell and the pitch angle $q$ between the A momentum and the 
sum of the B momenta, associated with the spiral order. The classical fields are accordingly written as:
\begin{eqnarray}
\mathbf{A}_l&=&\cos[q(2l-1)]\mathbf{x}+\sin[q(2l-1)]\mathbf{z};\nonumber\\
\mathbf{B}_{1l}&=&\cos(\theta)\cos(2ql)\mathbf{x}+\sin(\theta)(-1)^l\mathbf{y}\nonumber\\
               & &+\cos(\theta)\sin(2ql)\mathbf{z}, \text{and}\nonumber\\
\mathbf{B}_{2l}&=&\cos(\theta)\cos(2ql)\mathbf{x}+\sin(\theta)(-1)^{l+1}\mathbf{y}\nonumber\\
               & &+\cos(\theta)\sin(2ql)\mathbf{z}, 
\end{eqnarray}
with $|\mathbf{A}_l|=|\mathbf{B}_{1l}|=|\mathbf{B}_{2l}|\equiv 1$, while $\mathbf{x},\mathbf{y}~\text{and}~\mathbf{z}$ are
orthogonal unit vectors in the three-dimensional space. Substituting these 
fields in the classical  version of the Hamiltonian, \mbox{Eq. (\ref{Hr})}, we get the energy function 
$E(q,\theta) \sim 4{\cos}q\cos\theta + J(\cos2\theta +\cos2q +2{\cos}^2\theta\cos2q - 2{\sin}^2\theta)$
and minimizing this function with respect to $q$ and $\theta$, we find that $\cos(\theta)=1$ and $\cos(q)=\pi$ 
for $0<J<(1/3)$, which is the classical version of the Lieb-Mattis phase  found for $0<J<0.34$ in the 
quantum Hamiltonian, for both quantum rotors (FERRI phase) and spin system (Phase F1 of Ref. 12). For $(1/3)<J<1$ we obtain
\begin{eqnarray}
\cos(\theta)&=&\sqrt{\frac{1-J}{2J}};\label{classical1}\\
\cos(q)&=&-\cos(\theta),\label{classical2}
\end{eqnarray} 
which may be seen as the classical version of the CANTED phase ($0.34 \lesssim J \lesssim 0.75$) and of the spiral 
phase ($0.445 \lesssim J \lesssim 0.91$) found in Ref. 12. This phase holds some similarities with the 
second phase observed for the quantum rotors in the first single-site approach of Sec. III.
Finally, for $J>1$ the three chains 
are antiferromagnetically ordered with the B momenta lying in the $y$ direction and the A momenta ordered in the 
$z$ direction, which is the classical analog of the decoupled phase observed for $J>0.75$ for the quantum rotors and 
for  $J>0.91$ for the spin system \cite{Rene}. Such phase does not exist in the single-site approach: it is only obtained 
asymptotically ($J \rightarrow \infty$). Therefore, the classical solution presents two critical 
points: $J_{c1,\text{classical}}=1/3$ and $J_{c2,\text{classical}}=1$, and the transitions remain second order. Also, 
the first-order transition at $J=0.36$ ($J_t\approx 0.445$, for the spin systems) is not observed in the classical 
model. In fact, in the F$_2$ phase\cite{Rene} ($0.34\lesssim J\lesssim 0.445$) the number of singlets is quantized and the 
spiral peak is absent, while in the classical model the two orders coexist for 
$J_{c1,\text{classical}}<J<J_{c2,\text{classical}}$. This classical result is also indicated in \mbox{Fig. \ref{Figure15} (a)}.

\par In \mbox{Fig. \ref{Figure15} (b)}, we present the results of this classical interpretation for the momentum correlations.
A direct relationship with \mbox{Fig. \ref{Figure11}} can be established: we have the classical counterparts of the FERRI phase
($J \leq 1/3$) and the AF phase ($J\geq 1$); the CANTED phase is but a gradual continuous 
transition between the FERRI and AF phases. Further, the decoupling transition in the classical model is clearly 
observed at $J=1$ through the dot products indicated in the figure. Finally, we notice that when this classical approach is applied to frustration 
F$_1$, the minimum energy configuration obtained is the same as that derived through the first classical model discussed in 
Sec. III.

\section{Summary and Conclusions}

In dealing with quantum rotors placed at the sites of an AB$_2$ chain we resorted to a cluster variational MF
theory implemented via two distinct approaches, which yielded different results. First, we learned that the size of
the Hilbert space could be considerably reduced without affecting results, with the proviso that the rotor states should be 
kept possibly nearest ($\ell = 1/2$)-momentum states. That was attained by increasing the importance of the kinetic energy term in
the Hamiltonian (by setting relatively high values of the coupling $g$), whenever needed. This was a most important
fact for the computations in multiple-cell clusters in the second approach.
\par In the first approach the natural single-site MF theory was developed. A two-phase pattern was produced 
with the phase transition between them being of second order for $M = 0$ and of first order for $M =1$. For
$J \leq 1$, the Lieb-Mattis-like phase typical in the spin-1/2 system arose, and for $J \gg 1$, we observed the 
decoupling of the system where the momenta on the A sites tend to become uncorrelated with the momenta on the B sites,
which in turn formed singlet-like pairs, while the decoupling of the A sites was a salient feature laid bare by this approach, 
much like the dimer-monomer phase in Ref. 12. A classical interpretation was laid down that conformed to our 
QR numerical findings, inclusively showing how the fixed coupling momentum-position turns the second-order 
transition into a first-order one, when $M=1$. However, we were not able 
to provide a reasonable quantum picture that could relate to the known behavior of the corresponding quantum 
AB$_2$ chain. Furthermore, treated this way the system was not able to essentially tell apart frustration F$_1$ 
from frustration F$_2$ and this alone constituted a major setback. So, all this was a reminder that the main goal 
of our work still remained to be achieved.
\par In our last step, we then improved the former approach by producing a cluster (double-cell) variational MF theory in
which the trial Hamiltonian acts on the space composed of the tensor product of the respective local subspaces of
the six sites at two neighboring unit cells. The gist of this theory stands on the important fact that it allows the construction 
of the two-point correlations $<\hat{\mathbf{X}_i} \cdot \hat{\mathbf{Y}_j} >$ between any pairs of operators acting on sites 
$i,j$ of the cluster. This afforded us the observation of quantum features inherent in the system, as well as to distinguish 
between both frustrations F$_1$ and F$_2$. For the construction of this more complex ``system'', we
relied on the availability of processing capacity to carry out the numerical implementation.
\par For frustration F$_1$, besides the QR numerical simulation, we carried out ED on the spin-1/2 diamond chain using a system 
with 28 sites and calculated the correlation functions between spin at a central cluster, as well as other relevant 
physical quantities. Upon confronting with the QR results, we verified 
that the QR phase diagrams obtained through numerical implementation of the double-cell MF variational approach 
exhibited a sequence of phases analogous to those of the spin chains, with phase transitions of the same nature. We therefore
produced the FERRI-TD-DM phase sequence, with first order transitions, which is in essence the phase diagram of Ref. 10.
The transition point FERRI-TD at $J=0.68$ is somewhat displaced, but the transition point \mbox{TD-DM} at $J = 2$ was exactly
calculated by our approach.  
\par For frustration F$_2$,  we obtained a phase diagram in good agreement with the results of Ref. 12 on the 
respective spin-1/2 chains, endowed with the equivalent frustration pattern: FERRI, CANTED, and AF which are associated 
with the phases F1, F2/Spiral Singlet, and decoupled ladder chain, respectively, of the spin model. 
Notwithstanding, the criticality of the A spins correlations manifests itself here as an AF magnetic ordering
due to finite-size effects. For the same reason, we can not probe the short-rangedness of the 
correlation functions between B momenta. We also produced  ED as well as 
DMRG results that helped us to visualize the spiral phase in the spin system, 
and derived an insightful classical interpretation. 

\section{Acknowledgments}

This work was supported by CNPq, FACEPE, CAPES, and Finep (Brazilian agencies).

\appendix

\section{The Basis of Monopole Harmonics States}
In this appendix we provide the derivation of the matrix elements in the 
monopole harmonics basis representation for the $\hat{\bf{n}}$ operator, \textit{for any value of} $q$. 
This derivation was done straightforwardly based solely on the definitions \cite{Mono} and the 
properties of the Jacobi polynomials. Thus the following recurrence relations of the Jacobi 
polynomials \cite{Bate} can be established, valid  for all $q$:
\begin{widetext}
     \begin{eqnarray}
  l \left[\frac{(l+m+1)(l-m+1)(l+q+1)(l-q+1)}{(2l+3)(l+m)(l-m)} \right]^{1/2}Y_{q,l+1,m}(\theta,\phi) & = & \nonumber \\
  \left[l(l+1)\cos\theta + 2mq\right]\left[\frac{2l+1}{(l+m)(l-m)}\right]^{1/2}Y_{q,l,m}(\theta,\phi) &   & \nonumber \\
  - (l+1)\left[\frac{(l+q)(l-q)}{2l-1}\right]^{1/2}Y_{q,l-1,m}(\theta,\phi), &  & \label{me1}
                                   \end{eqnarray}
                                  \begin{eqnarray}
\hspace{0.75in}\left[(1-x^{2})e^{-i\phi}\right]\left[\frac{1}{(l+m)(l+m-1)}\right]^{1/2}Y_{q,l-1,m+1}(\theta,\phi) & = & 
\nonumber \\
\frac{1}{l}\left[\frac{(l+q)(l-q)(l-m-1)(l-m)}{(2l+1)(2l-1)(l+m)(l+m-1)}\right]^{1/2}Y_{q,l,m}(\theta,\phi) & & \nonumber \\
- \frac{q}{l(l-1)}\left[\frac{l-m-1}{l+m-1}\right]Y_{q,l-1,m}(\theta,\phi) &   & \nonumber \\
-\frac{1}{l-1}\left[\frac{(l+q-1)(l-q-1)}{(2l-3)(2l-1)}\right]^{1/2}Y_{q,l-2,m}(\theta,\phi), &   & \label{me2}
                                     \end{eqnarray}
                                        \begin{eqnarray}
\left[(1-x^{2})e^{i\phi}\right]\left[\frac{1}{(l-m+1)(l-m)}\right]^{1/2}Y_{q,l,m-1}(\theta,\phi) & = & \nonumber \\
-\frac{q(l+m)}{l(l+1)}\left[\frac{1}{(l+m)(l-m)}\right]^{1/2}Y_{q,l,m}(\theta,\phi) & & \nonumber \\
- \frac{1}{(l+1)}\left[\frac{(l+m)(l+m+1)(l+q+1)(l-q+1)}{(2l+1)(2l+3)(l-m+1)(l-m)}\right]^{1/2}Y_{q,l+1,m}(\theta,\phi) &   &
\nonumber \\
+\frac{1}{l}\left[\frac{(l+q)(l-q)}{(2l+1)(2l-1)}\right]^{1/2}Y_{q,l-1,m}(\theta,\phi). &  & \label{me3}
                                      \end{eqnarray}
\end{widetext}
Next we use the orthogonality relation for the monopole harmonics and specialize in the case $q=\frac{1}{2}$ to get 
the respective non-zero matrix elements for the operator $\bf\hat{n}$, used in this work. 
So, we have set $\hat{n}_{\pm} = \hat{n}_{x}\pm i\hat{n}_{y}$, which is defined similarly to $\hat{L}_{\pm}$, where we have the 
following: 
$x = \cos\theta$ and  $\hat{n}_{z} \leftrightarrow \cos\theta$, $\hat{n}_{+} \leftrightarrow e^{i\phi}\sin\theta$ (the matrix elements 
of $\hat{n}_{-}$ are obtained by complex conjugation):
$<\frac{1}{2},l,m|\hat{n}_{z}|\frac{1}{2},l,m> = -\frac{m}{l(2l+2)}$; 
$<\frac{1}{2},l,m|\hat{n}_{z}|\frac{1}{2},l-1,m> = \frac{1}{2l}\sqrt{(l-m)(l+m)}$;
\mbox{$<\frac{1}{2},l,m|\hat{n}_{+}|\frac{1}{2},l,m-1> = -\frac{1}{2l(l+1)}\sqrt{(l-m+1)(l+m)}$};
$<\frac{1}{2},l,m|\hat{n}_{+}|\frac{1}{2},l-1,m-1> = -\frac{1}{2l}\sqrt{(l+m)(l+m-1)}$; and
$<\frac{1}{2},l,m|\hat{n}_{+}|\frac{1}{2},l+1,m-1> = \frac{1}{2(l+1)}\sqrt{(l-m+1)(l-m+2)}$.
\par For the sake of completeness, we write down the matrix elements for the operator $\hat{\mathbf{L}}$, valid for all $q$,
obtained through the ladder-operator and eigenvalue relations for the monopole harmonics: \mbox{$<l,m|\hat{L}^2|l,m> = l(l+1)$}, \mbox{$<l,m|\hat{L}_{z}|l,m>= m$}, and \mbox{$<l,m|\hat{L}_{+}|l,m-1> = [(l+m)(l-m+1)]^{1/2}$}.


\begin{thebibliography}{30}
\bibitem{Auer} A. Auerbach, {\it Interacting Electrons and Quantum Magnetism} (Springer-Verlag, New York, 1998).
\bibitem{QPTsach} S. Sachdev, {\it Quantum Phase Transitions} (Cambridge University Press, Cambridge-UK, 2001).
\bibitem{HKS} C. J. Hamer, J. B. Kogut, and L. Susskind, Phys. Rev. D, {\bf 19}, 3091 (1979).
\bibitem{Haldane} F. D. M. Haldane, Phys. Lett. {\bf 93A}, 464 (1983); F. D. M. Haldane, Phys.Rev. Lett. {\bf 50}, 1153 (1983).
\bibitem{Shankar} R. Shankar and N. Read, Nucl. Phys. B {\bf 336}, 457 (1990).
\bibitem{Mono} T. T. Wu and C. N. Yang, Nucl. Phys. B {\bf 107}, 365 (1976).
\bibitem{ZPTsach} S. Sachdev and T. Senthil, Ann. Phys. {\bf 251}, 76 (1996).
\bibitem{Imambekov} A. Imambekov, M. Lukin, and E. Demler, Phys. Rev. Lett. {\bf 93}, 120405 (2004). 
\bibitem{Polak} T. P. Polak and T. K. Kope$\acute{c}$, Phys. Rev. B {\bf 76}, 094503 (2007).
\bibitem{TKS} K. Takano, K. Kubo, and H. Sakamoto, J. Phys.: Condens. Matter {\bf 8}, 6405 (1996).
\bibitem{TOHTK} T. Tonegawa, K. Okamoto, T. Hikihara,  Y. Takahashi, and M. Kaburagi, J. Phys. Soc. Jpn. {\bf 69}, 
                332 (2000).
\bibitem{Rene} R. R. Montenegro-Filho and M. D. Coutinho-Filho, Phys. Rev. B {\bf 78}, 014418 (2008).
\bibitem{frust1a} K. Sano and K. Takano, J. Phys. Soc. Jpn. {\bf 69}, 2710 (2000);
                  K. Okamoto, T. Tonegawa, Y. Takahashi, and M. Kaburagi, J. Phys.: Condens. Matter {\bf 11}, 
                  10485 (1999); H. Niggemann, G. Uimin, and J. Zittartz, J. Phys.: Condens. Matter {\bf 9}, 
                  9031 (1997).
\bibitem{frust1b} K. Okamoto, T. Tonegawa, and M. Kaburagi, J. Phys.: Condens. Matter {\bf 15}, 5979 (2003).
\bibitem{cyclic} N. B. Ivanov, J. Richter, and J. Schulenburg, Phys. Rev. B {\bf 79}, 104412 (2009).
\bibitem{Lanczos} C. Lanczos, J. Res. Natl. Bur. Stand. {\bf 45}, 255 (1950).
\bibitem{DMRG} S. R. White, Phys. Rev. B {\bf 48}, 10345 (1993); U. Schollw\"{o}ck, Rev. Mod. Phys. {\bf 77}, 259 (2005).
\bibitem{azurite} H. Kikuchi, Y. Fujii, M. Chiba, S. Mitsudo, T. Idehara, T. Tonegawa, K. Okamoto, T. Sakai, 
                  T. Kuwai, and H. Ohta, Phys. Rev. Lett. {\bf 94}, 227201 (2005); See also K. C. Rule, 
                  A. U. B. Wolter, S. S\"{u}llow, D. A. Tennant, A. Br\"{u}hl, S. K\"{o}hler, B. Wolf, M. Lang, and 
                  J. Schreuer, Phys. Rev. Lett. {\bf 100}, 117202 (2008).
\bibitem{Matsuda} M. Matsuda, K. Kakurai, A. A. Belik, M. Azuma, M. Takano, and M. Fujita, 
                  Phys. Rev. B {\bf 71}, 144411 (2005).
\bibitem{Hosokoshi} Y. Hosokoshi, K. Katoh, Y. Nakazawa, H. Nakano,  and K. Inoue, J. Am. Chem. Soc. {\bf 123}, 
                    7921 (2001).
\bibitem{revisao} For a review see: M. D. Coutinho-Filho, R. R. Montenegro-Filho, E. P. Raposo, C. Vitoriano, and M. H.Oliveira, 
                  J. Braz. Chem. Soc. {\bf 19}, 232 (2008).
\bibitem{Hubb} A. M. S. Mac\^{e}do, M. C. dos Santos, M. D. Coutinho-Filho, and C. A. Mac\^{e}do, Phys. Rev. Lett. {\bf 74}, 
               1851 (1995); G.-S. Tian and T.-H. Lin, Phys. Rev. B {\bf 53}, 8196 (1996).
\bibitem{tJ} G. Sierra, M. A. Mart\'{\i}n-Delgado, S. R. White, D. J. Scalapino and J. Dukelsky, 
             Phys. Rev. B {\bf 59}, 7973 (1999)
\bibitem{ClHeis_I} C. Vitoriano, M. D. Coutinho-Filho, and E. P. Raposo, J. Phys. A: Math. Gen. {\bf 35}, 9049 (2002);
\bibitem{Heis} F. C. Alcaraz and A. L. Malvezzi, J. Phys. A: Math. Gen.  {\bf 30}, 767 (1997); E. P. Raposo and 
               M. D. Coutinho-Filho, Phys. Rev. Lett. {\bf 78}, 4853 (1997); E. P. Raposo and 
               M. D. Coutinho-Filho, Phys. Rev. B {\bf 59}, 14384 (1999); 
               M. A. Mart\'{\i}n-Delgado, J. Rodriguez-Laguna, and G. Sierra, Phys. Rev. B   {\bf 72}, 104435 (2005).
\bibitem{magnexc1} C. Vitoriano, F. B. de Brito, E. P. Raposo, and M. D. Coutinho-Filho, 
                   {\it Mol. Cryst. Liq. Cryst.} {\bf 374}, 185 (2002); T. Nakanishi and S. Yamamoto, 
                   Phys. Rev. B {\bf 65}, 214418 (2002); S. Yamamoto and J. Ohara, Phys. Rev. B {\bf 76}, 
                   014409 (2007) 
\bibitem{magnexc2} R. R. Montenegro-Filho and M. D. Coutinho-Filho, Physica A {\bf 357}, 173 (2005).
\bibitem{Mario} M. H. Oliveira, M. D. Coutinho-Filho, and E. P. Raposo, Phys. Rev. B {\bf 72}, 214420 (2005). 
\bibitem{hole} R. R. Montenegro-Filho and M. D. Coutinho-Filho, Phys. Rev. B {\bf 74}, 125117 (2006), and 
               references therein.
\bibitem{Call} H. B. Callen, {\it Thermodynamics and an Introduction to Thermostatistics} 
               (John Wiley \& Sons Inc, New York-NY, 1985); R. P. Feynman, {\it Statistical Mechanics--A Set 
               of Lectures} (The Benjamin Cummings Publishing Company Inc, Reading-MA, 1972).
\bibitem{Yokoi} C. S. O. Yokoi, M. D. Coutinho-Filho, and S. R. Salinas, Phys. Rev. B {\bf 24}, 4047 (1981). 
\bibitem{Simplex} {\it Numerical Recipes Inc: The Art of Scientific Computing} (Cambridge University 
                  Press, Cambridge-UK, 1992).
\bibitem{Lieb} E. H. Lieb and D. Mattis, J. Math. Phys. {\bf 3}, 749 (1962). 
\bibitem{Bate} A. Ederlyi (editor), {\it Higher Transcendental Functions (Bateman Project)} (McGraw--Hill Book
        Company Inc, New York-NY, 1953).



\end{thebibliography}
\end{document}